\documentclass[prd,aps,amsfonts,superscriptaddress,nofootinbib,longbibliography,notitlepage,twocolumn]{revtex4-1}

\usepackage{graphicx}
\usepackage{xcolor}
\usepackage{rotating}
\usepackage{amsmath,amssymb,graphics,amsthm,isomath}
\usepackage{bm}

\usepackage[colorlinks=true, urlcolor=violet, linkcolor=blue, citecolor=red, hyperindex=true, linktocpage=true]{hyperref}
\usepackage[capitalise,compress]{cleveref}

\makeatletter
\renewcommand{\p@subsection}{}
\renewcommand{\p@subsubsection}{}
\makeatother

\usepackage{xcolor}
\usepackage{mathtools}

\newcommand{\ii}{\mathrm{i}}

\newcommand{\PP}{\mathbb{P}}
\newcommand{\LL}{\mathcal{L}}

\newcommand{\slr}[1]{\left( #1\right)}
\newcommand{\mlr}[1]{\left[ #1\right]}
\newcommand{\norm}[1]{\left\lVert#1\right\rVert}

\newcommand{\comment}[1]{}
\usepackage{dsfont}

\begin{document}
\title{Quantum operator growth bounds for kicked tops and semiclassical spin chains}

\author{Chao Yin}
\email{chao.yin@colorado.edu}
\affiliation{Department of Physics and Center for Theory of Quantum Matter, University of Colorado, Boulder, CO 80309, USA}

\author{Andrew Lucas}
\email{andrew.j.lucas@colorado.edu}
\affiliation{Department of Physics and Center for Theory of Quantum Matter, University of Colorado, Boulder, CO 80309, USA}

\begin{abstract}
We present a framework for understanding the dynamics of operator size, and bounding the growth of out-of-time-ordered correlators, in models of large-$S$ spins. Focusing on the dynamics of a single spin, we show the finiteness of the Lyapunov exponent in the large-$S$ limit; our bounds are tighter than the best known Lieb-Robinson-type bounds on these systems.  We numerically find our upper bound on Lyapunov exponents is within an order of magnitude of numerically computed values in classical and quantum kicked top models. Generalizing our results to coupled large-$S$ spins on lattices, we show that the butterfly velocity, which characterizes the spatial speed of quantum information scrambling, is finite as $S\rightarrow\infty$.   We emphasize qualitative differences between operator growth in semiclassical large-spin models, and quantum holographic systems including the Sachdev-Ye-Kitaev model.

\end{abstract}

\date{\today}

\maketitle

\section{Introduction}
 Quantum systems with a natural semiclassical limit are the canonical setting for understanding quantum chaos, and its connections with classical physics \cite{haake87,haake92,kus1993quantum,CHIRIKOV198877,aleiner,agam}.  Several years ago, this subject was revived by the study of out-of-time-ordered correlators (OTOC) in \emph{many-body} systems \cite{larkin,Shenker:2013pqa,Roberts:2014isa,Stanford:2015owe,Maldacena:2015waa,Patel:2016wdy,werman2017quantum}.  These OTOCs grow exponentially in various large-$N$ quantum systems with non-local interactions, mimicking the rapid deviation of trajectories in a classically chaotic system: \begin{equation}
    -\left\langle [\mathcal{O}_1(t), \mathcal{O}_2]^2\right\rangle \sim N^{-1} \mathrm{e}^{\lambda t}.  \label{eq:introotoc}
\end{equation} 
Here $\mathcal{O}_{1,2}$ correspond to ``small" local operators. This exponential growth of OTOCs can, in some circumstances, be deeply related to holographic quantum gravity \cite{Susskind:2018tei,Roberts:2018mnp,Qi:2018bje,Mousatov:2019xmc,Magan:2020iac,Lensky:2020ubw}: the growth of an ``operator size distribution" has been conjectured to probe the emergent holographic dimension of quantum gravity.  

Motivated by this holographic connection, a significant literature has arisen \cite{Chew_2017,Bentsen_2019,Chen_2018,Marino_2019,Lewis_Swan_2019,Alavirad_2019,Bentsen_2019prl,belyansky2020minimal,li2020fast}, aiming to construct experimentally simulatable models of many-body quantum chaos and quantum gravity.  Some of this theoretical literature focuses on large-$S$ spin models \cite{Scaffidi:2017ghs, Pappalardi:2019kkh,Bentsen_2019prx,Sieberer_2019,PhysRevA.102.032404}, because the $1/S$ expansion behaves much like the $\hbar$ expansion \cite{haake87}, and so classically chaotic Hamiltonian systems should become quantum chaotic ones.  However, it is, as of yet, unclear how operators grow in such models, and even whether questions of operator growth are well defined, let alone related to quantum gravity.

In this paper, we present the necessary mathematical framework to make precise statements about operator growth and OTOC dynamics in large-$S$ spin models.  We provide a natural definition of operator size and operator growth in such a model, despite the lack of a genuinely many-body Hilbert space.   We then use recently developed ``operator quantum walk" methods \cite{lucas2020non,Tran:2020xpc,Yin:2020pjd,Lucas:2020pgj} to prove that the growth exponent $\lambda$ of the OTOC does not depend on $S$.  In large-$S$ spin chains and lattices, a straightforward generalization of our framework proves that (\emph{1}) the butterfly velocity $v_{\mathrm{B}}$, characterizing the spatial growth of the OTOC \cite{Roberts:2014isa,das18lightcone,nahum_operator_2018,von_keyserlingk_operator_2018}, is also independent of $S$, and (\emph{2}) the prefactor of the OTOC is suppressed by $1/S$, as anticipated from (\ref{eq:introotoc}).

The Lieb-Robinson theorem \cite{Lieb1972,Hastings:2005pr} has long been the canonical method of choice for proving locality and constraining quantum information dynamics.  Using recently developed methods \cite{PRXQuantum.1.010303}, it is possible to show that both $\lambda$ and $v_{\mathrm{B}}$ are independent of $S$. However, we show that for general models, our framework based on operator size growth leads to tighter bounds on both $\lambda$ and $v_{\mathrm{B}}$.  In fact, our bounds on $\lambda$ are within an order of magnitude of numerically calculated exponents in semiclassical kicked top models, demonstrating that our bounds are strong enough to give non-trivial limits on \emph{the classical limit} of quantum dynamics.

Our framework also gives a more rigorous perspective on subtle, but important, differences in OTOC growth and scrambling between many-body chaotic models with and without a semiclassical limit.   While in a holographic model the operators which grow rapidly and dominate the OTOC are exponentially rare (among large operators) \cite{lucas2020non},  typical operators in semiclassical models can grow rapidly. Our techniques thus help shed light on which experimentally realizable models genuinely mimic scrambling in quantum holographic models, and which ones do not.

\section{Results}

 We now introduce the models of interest.  Let the spin operators for a spin-$S$ degree of freedom be  $\bm{S} = (S^x, S^y, S^z)= (X,Y,Z)$, namely $[S^\alpha, S^\beta]=\ii \epsilon^{\alpha\beta\gamma}S^\gamma, \bm{S}^2=\sum_\alpha (S^\alpha)^2=S(S+1)$.  We first consider a general time-dependent Hamiltonian
\begin{equation}\label{eq:generalH}
    H(t) = \sum_{\bm{n}} h(\bm n;t)S^{1-n_x-n_y-n_z} X^{n_x}Y^{n_y}Z^{n_z},
\end{equation}
where $\bm{n}\equiv (n_x, n_y, n_z)$. The terms are normalized by powers of $S$ to have a well-defined semiclassical limit, while $h(\bm n; t)$ is independent of $S$. We define a canonical infinite temperature OTOC for an initial operator $\mathcal{O}$: \begin{equation}
    F(\mathcal{O}(t))=-\frac{1}{2S+1}\sum_\alpha \mathrm{tr}[S^\alpha, \mathcal{O}(t)]^2. \label{eq:fotoc}
\end{equation}
As we will see, by decomposing $\mathcal{O}(t)$ into irreducible tensors of the rotation group $\mathrm{SU}(2)$, we can interpret $F(\mathcal{O}(t))$ as the average size of $\mathcal{O}(t)$.

In a chaotic model, $F$ can grow exponentially: $F(\mathcal{O}(t))\sim \exp[\lambda_{\mathrm{OTOC}}t]$ for times $t\lesssim \ln S$.  Our first main result is that this is, in fact, the \emph{fastest} possible OTOC growth: $\lambda_{\mathrm{OTOC}}$ is bounded by a constant independent of $S$.  A general proof is given in Appendix \ref{app:proofs}.
\comment{

}
In the main text, we focus our discussion on a well-studied semiclassical spin model: the kicked top \cite{haake87}
\begin{equation}
    H(t)=\frac{\kappa}{2S+1}Z^2+hX\sum_{n\in\mathbb{Z}}\delta(t-n). \label{eq:kickedrotor}
\end{equation}
Here, we will prove that $\lambda_{\mathrm{OTOC}}\le \kappa$; moreover, by studying the dynamics numerically, we find $\lambda_{\mathrm{OTOC}} \le 0.44\kappa$, which is within an order of magnitude of our rigorous constraint.  


We then turn to models with spatial structure.  As a simple example, consider a chain of coupled (kicked) tops \begin{equation}\label{eq:Hchain}
    H = \sum_{i=1}^L  \left[h_i(t)X_i + \frac{1}{2S+1} \left(\kappa'Z_i^2 +\kappa Z_i Z_{i+1}\right) \right]
\end{equation}
where $L$ is arbitrarily large and denotes the size of the system.  Our definition of operator size immediately generalizes to this large-$S$ system, and we prove that $\lambda$ and $v_{\mathrm{B}}$ are finite. In each of the two cases above, our results are generically stronger than the best known Lieb-Robinson bounds (when $\kappa'\ne 0$).

\section{Operator growth}
Let us now develop the operator growth formalism and why $\lambda_{\mathrm{OTOC}}\le\kappa$ in the kicked top model.  The first step is a suitable definition of operator size, the average of which relates to the OTOC.  Observe that if we start with a single spin $S$ Hilbert space, which transforms in the representation $S$ of SU(2), that the vector space of all operators contains representations \begin{equation}
    S\otimes S = 0 \oplus 1 \oplus \cdots \oplus 2S.
\end{equation} 
The ``spherical-harmonic tensor operators" $\{\mathcal{Y}^m_l: m=-l,\cdots,l; l=0,\cdots,2S\}$ \cite{Li_2013} are irreducible tensors transforming in representation $l$:  \begin{subequations}\label{eq:comm}\begin{align}
    [Z, \mathcal{Y}^m_l]&=m\mathcal{Y}^m_l, \\
    [X\pm\ii Y, \mathcal{Y}^m_l] &= \sqrt{(l\mp m)(l\pm m+1)} \mathcal{Y}^{m\pm 1}_l.
\end{align}\end{subequations}
They are, intuitively, the operator generalizations of the well-known spherical harmonics.  The operators $\mathcal{Y}_l^m$ can be found by writing $r^l Y_l^m(\theta,\phi)$ in rectangular coordinates, and subsequently replacing products such as $xy$ with $\frac{1}{2}\lbrace X,Y\rbrace$, etc. More alegbraically, we start with \begin{equation}
    \mathcal{Y}^{\pm l}_l = \frac{(\mp1)^l}{2^ll!}\sqrt{\frac{(2l+1)!}{4\pi}} (X\pm \mathrm{i} Y)^l,
\end{equation}
from which we can use the su(2) algebra (\ref{eq:comm}) to find all $\mathcal{Y}_l^m$.  All $\mathcal{Y}_l^m$ are orthogonal to each other. 

The linearity of quantum mechanics implies that operators themselves live in a ``Hilbert space".  We often write an operator $\mathcal{O}$ in the bra-ket-like notation $|\mathcal{O})$ to emphasize this fact.  Defining the operator inner product \begin{equation}
    (A|B) = \frac{1}{2S+1}\mathrm{tr}(A^\dagger B),
\end{equation} we find that
\begin{align}\label{eq:Y_norm}
    (\mathcal{Y}^m_l|\mathcal{Y}^m_l) =\frac{1}{4\pi}\prod^l_{k=1}\mlr{(S+\frac{1}{2})^2-\frac{1}{4}k^2}.
\end{align}
We define the normalized basis $T^{lm}=\mathcal{Y}^m_l/\norm{\mathcal{Y}^m_l}$.  
 Each operator can be expanded in this basis: \begin{equation}
    \mathcal{O}(t)=\sum_{l,m} O_{lm}(t) T^{lm}.
\end{equation}
Lastly, we define the following projection operators: $\mathbb{Q}_{l^\prime}|T^{lm}) =\delta_{ll^\prime} |T^{lm})$.

By unitarity, \begin{equation}
  \frac{\mathrm{d}}{\mathrm{d}t} \sum_{l,m}  |\mathcal{O}_{lm}|^2 = 0.
\end{equation}
Hence we can define a probability distribution of ``operator size":  an operator has size $l$ with probability $\phi_l^2$: \begin{equation}
    \phi_l^2 = \sum_{m=-l}^l \frac{|\mathcal{O}_{lm}|^2}{(\mathcal{O}|\mathcal{O})}
\end{equation}
Normalizing $(\mathcal{O}|\mathcal{O})=1$ and using (\ref{eq:comm}), the OTOC (\ref{eq:fotoc}) is \begin{equation}
    F(\mathcal{O})=\sum_{l=0}^{2S} l(l+1) \phi_l^2.
\end{equation} 

Rather than bound the OTOC directly a la Lieb-Robinson, we now bound OTOCs by constraining the ``stochastic process" governing the time evolution of the operator size distribution.  Because operators evolve in time according to \begin{equation}
    \frac{\mathrm{d}}{\mathrm{d}t} |\mathcal{O}(t)) = \mathcal{L}|\mathcal{O}(t)) = |\mathrm{i}[H,\mathcal{O}(t)]),
\end{equation} we can formally write down a set of linear equations for the probability amplitudes $\phi_l$:  \cite{lucas2020non,Tran:2020xpc,Yin:2020pjd,Lucas:2020pgj}
\begin{equation}\label{eq:qwalk}
    \frac{\mathrm{d}\phi_l}{\mathrm{d}t} = K_{l-1}(t)\phi_{l-1}(t) - K_{l}(t)\phi_{l+1}(t),
\end{equation}
where the coefficients $K_l(t)$ obey \begin{equation}
    K_l(t) \le \lVert \mathbb{Q}_l \mathcal{L} \mathbb{Q}_{l+1}\rVert \equiv \mathcal{K}_l,
\end{equation}
where $\lVert \cdots \rVert$ denotes the conventional operator norm, acting on the space of superoperators.



To bound $\mathcal{K}_l$, notice that the only size changing terms arise from $\mathrm{i}[Z^2,T^{lm}] = \mathrm{i}m \lbrace Z,T^{lm}\rbrace$.  Now observe that 
\begin{align}\label{eq:TT}
    \mathcal{Y}^{0}_1T^{lm} = \sum_{l'=l-1}^{l+1}  c(l,l') \langle 10lm|l' m\rangle T^{l' m},
\end{align}
where $Z=2\sqrt{\pi/3}\mathcal{Y}^{0}_1$, $\langle 1m'lm|l' (m+m')\rangle$ is the Clebsch-Gordan coefficient, and $c(l,l')$ is a constant independent of $m,m'$ related to normalization of $T^{lm}$.  An explicit calculation gives
\begin{equation}
    c(l,l+1)=\frac{1}{2}\sqrt{\frac{3(l+1)}{\pi(2l+3)}}\sqrt{(S+\frac{1}{2})^2-\frac{1}{4}(l+1)^2}.
\end{equation}
We can similarly evaluate $T^{lm}\mathcal{Y}^{0}_1$, and eventually find
\begin{equation}
 \mathbb{Q}_l \mathcal{L}T^{l-1,m} = \mathrm{i}\kappa m \sqrt{\frac{[l^2-m^2][(2S+1)^2-l^2]}{(2l-1)(2l+1)(2S+1)^2}}T^{lm} \label{eq:Qlcoefficient}
\end{equation}
A crucial observation is that, $\mathbb{Q}_{l+1}\LL T^{lm}$ and $\mathbb{Q}_{l+1}\LL T^{lm^\prime}$ are orthogonal if $m\ne m^\prime$.  Hence $\mathcal{K}_l$ can be upper bounded by the maximal value of the coefficient in (\ref{eq:Qlcoefficient}): \begin{equation}\label{eq:Kl<}
    \mathcal{K}_l= \kappa\frac{\max_m m\sqrt{(l+1)^2-m^2} }{\sqrt{(2l+1)(2l+3)}} \le \frac{\kappa (l+1)^2}{2\sqrt{(2l+1)(2l+3)}}.
\end{equation} 
We then explicitly find that \cite{lucas2020non} 
\begin{align}
    \frac{\mathrm{d}F}{\mathrm{d}t} &= 2\sum_{l=0}^{2S} l(l+1)\phi_l \frac{\mathrm{d}\phi_l}{\mathrm{d}t} 
    \le \sum_{l=0}^{2S} 2(l +1)\mathcal{K}_l \left(\phi_l^2 + \phi_{l+1}^2\right) \notag \\
    &\le \kappa \sum_{l=0}^{2S} (l(l+1)+2)\phi_l^2 = \kappa (F+2).
\end{align}
Thus for kicked top models, \begin{equation}
    \lambda_{\mathrm{OTOC}}\le \kappa. \label{eq:lotocbound}
\end{equation} 

Our bound (\ref{eq:lotocbound}) is not saturated without the ``kicks".  If $h=0$ in (\ref{eq:kickedrotor}), $H\propto Z^2$, $T^{11}(t)$ only hops on sites $\{T^{l1}\}$, and $|K_l(t)| \le \frac{1}{2}$. As a result, $F(t) \le \mathit{O}(t^2)$. Although this is expected because (\ref{eq:kickedrotor}) is integrable when $h=0$, our formalism crisply captures how this integrability prevents the operator from growing rapidly.  Kicks, which move us from operators with $m\sim 1$ to $m\sim l$, are required to come close to saturating our bound.

\section{Classical and Quantum Kicked Top} 
\begin{figure}[t]
\includegraphics[width=0.5\textwidth]{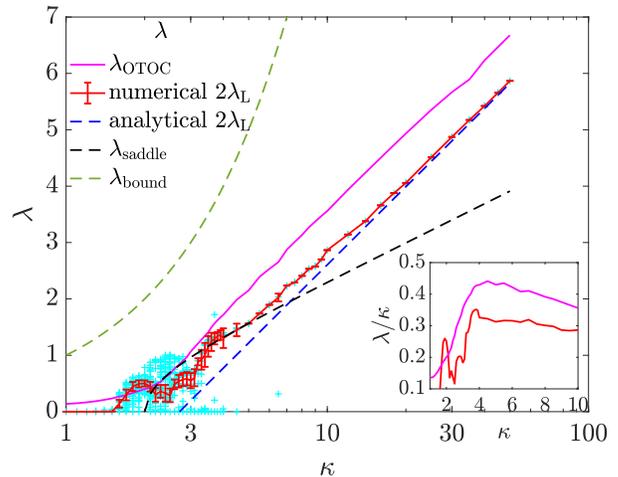}
\caption{\label{fig:LE_KT} 
Classical and quantum LE for kicked top at $h=\pi/2$. The main plot: The magenta line is the quantum LE $\lambda_{\mathrm{OTOC}}$. The red line is classical LE $2\lambda_{\mathrm{L}}$ averaged for $100$ initial states, with the standard deviation as errorbar. The cyan scattered points are for each initial points. 
The blue dashed line is the analytical result $\lambda_{\mathrm{L}} =\ln\slr{\kappa |\sin h|}-1$. The black dashed line is $\lambda_{\mathrm{saddle}}=\ln\slr{\frac{\kappa}{2}+ \sqrt{\kappa^2/4-1}}$. The green dashed line is our bound $\lambda_{\mathrm{bound}}=\kappa$.  The inset is of the same data for $\lambda_{\mathrm{OTOC}}$ and numerical $2\lambda_{\mathrm{L}}$ in the main plot, with axes modified. }
\end{figure}

We now compare (\ref{eq:lotocbound}) to the actual value of $\lambda_{\mathrm{OTOC}}$ in the semiclassical kicked top.  When comparing to the standard definition of Lyapunov exponent (LE), we must replace $2\lambda_{\mathrm{L}} = \lambda_{\mathrm{OTOC}}$.  When $\kappa$ is large, one finds that  \cite{ktop97} \begin{equation}
    \lambda_{\mathrm{L}} =\ln\slr{\kappa |\sin h|}-1.
\end{equation}
At finite $\kappa $, we have numerically calculated the classical LE by the tangent map matrix method \cite{ktop97}, evolving for $0<t<10^6$.  We set $h=\frac{\pi}{2}$ to optimize operator growth.  The result is shown in Fig.~\ref{fig:LE_KT}, in comparison to our fully quantum mechanical bound.

\begin{figure*}
\includegraphics[width=\textwidth]{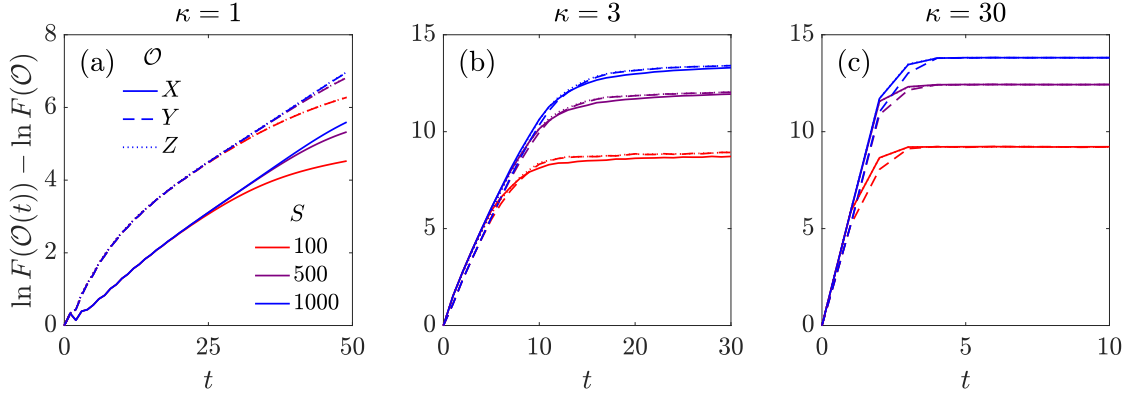}
\caption{\label{fig:otoc} Dynamics of the OTOC in the quantum kicked top at $h=\pi/2$.  The three line types correspond to three initial operators $\mathcal{O}=X,Y,Z$. The three colors/shades correspond to $S=100,500,1000$, as shown in the legend.  A clear exponential behavior is observed in this plot of $\ln F$; note however that the late time LE (given by the slope at larger $t$) is always smaller than $\ln F(1) - \ln F(0)$ -- the increase in the (log) OTOC after the first time step. }
\label{fig:data}
\end{figure*}

We have also, for smaller system sizes, calculated the quantum OTOCs of a kicked top: by calculating $F(\mathcal{O}(t))$ numerically, we extract the quantum LE by fitting an exponential growth in time at $S=500$, while averaging the exponents over initial operator $\mathcal{O}=X,Y,Z$.  One finds the quantum LE is indeed larger than the classical one, and at large $\kappa$ the difference goes to a constant $\approx 0.4$, which qualitatively agrees with that in the kicked rotor \cite{Victor17}. The largest $\lambda_{\mathrm{OTOC}}/\kappa$ is $0.44$ at $\kappa\approx 4.5$, while the largest $2\lambda_{\mathrm{L}}/\kappa$ is $0.35$ at $\kappa\approx 3.7$, as shown in the inset. These are smaller than our bound, but are well within an order of magnitude. At large $\kappa$ the LEs $\propto \ln\kappa $ are parametrically smaller than our bound, which is consistent with the fact that $\lambda_{\mathrm{OTOC}}$ is always smaller than the exponent $\ln F(\mathcal{O}(1))- \ln F(\mathcal{O})$ in the first period: see Figure \ref{fig:data}. The operator growth during that first time period is induced solely from the $Z^2$ term in $H$, where $F(\mathcal{O}(t))-F(\mathcal{O})\sim c_1(\kappa t)^2+c_2\kappa t$ with constant $c_1, c_2$.  This leads to $\lambda_{\mathrm{OTOC}}\le 2\ln\kappa$ at large $\kappa$, although this behavior is not captured by (\ref{eq:lotocbound}) which allows for arbitrarily strong kicking.

When $\kappa \ge 2$, following \cite{PhysRevB.98.134303,scram_not_chaos20}, we can also argue for a lower bound \begin{equation}\label{eq:saddle}
    \lambda_{\mathrm{OTOC}}\ge \lambda_{\mathrm{saddle}}=\ln\slr{\frac{\kappa}{2}+\sqrt{\frac{\kappa^2}{4}-1}}
\end{equation}
Here $\lambda_{\mathrm{saddle}}$ refers to the growth rate of OTOCs arising from an unstable point in the semiclassical ``phase space" as follows. We start from the classical limit of (\ref{eq:kickedrotor}), which leads to the dynamical map \cite{ktop97} for $h=\pi/2$:
\begin{align}
    J^z_{t+1} &= J^y_t,\nonumber\\
    J^x_{t+1} &= J^x_t \cos(\kappa J^z_{t+1})+ J^z_t \sin(\kappa J^z_{t+1}), \nonumber\\
    J^y_{t+1} &= J^x_t \sin(\kappa J^z_{t+1})- J^z_t \cos(\kappa J^z_{t+1}),
\end{align}
where $\bm{J}=(J^x, J^y, J^z)$ is a classical vector with norm $1$. One can verify that $\bm{J}=(\pm1,0,0)$ are $2$ fixed points. Linearize the map near $(1,0,0)$ for example (the other fixed point has the same result):
\begin{align}
    J^z_{t+1} &= J^y_t, \nonumber\\
    J^y_{t+1} &= \kappa J^z_{t+1}-J^z_t.
\end{align}
Using the ansatz $J^{z,y}_t \sim \delta J_{z,y}^0 \times \mathrm{e}^{\omega t}$, we solve for $\omega$ and find the ``quasinormal modes" \begin{equation}
     \mathrm{e}^{\omega_\pm} = \frac{\kappa \pm\sqrt{\kappa^2 -4}}{2}.
\end{equation} Thus we conclude there is an unstable sadddle point when $\kappa>2$, and following \cite{scram_not_chaos20}, we predict $\lambda_{\mathrm{saddle}}=\omega_+$ in eq.(\ref{eq:saddle}). Note that there are other nontrivial saddle points in the classical kicked top \citep{kus1993quantum}, which might give a better bound. Hence, this saddle point physics is as important as semiclassical chaos when saturating our bound, since $\lambda_{\mathrm{saddle}}$ is comparable to $\lambda_{\mathrm{L}}$ in the region around $\kappa\approx 4$, where our bound is tightest. 

\section{Semiclassical Spin Chains} Next, let us discuss operator growth in the interacting semiclassical spin chain eq.(\ref{eq:Hchain}), where operators are characterized by their size $l$ on each site. Define $\mathbb{P}_{il}$ as the projector onto operators which have size $l$ on site $i$.  (Note that for $i\ne i^\prime$, $\mathbb{P}_{il}$ and $\mathbb{P}_{i^\prime l^\prime}$ do not project  onto disjoint sets.)
Consider an initial operator $\mathcal{O}$ which has support only on vertex $i_0=1 \in V$, and obeys $(\mathcal{O}|\mathcal{O})=1$. Using similar methods to our earlier discussion, we prove a bound on how fast $\mathcal{O}(t)$ can grow:
\begin{equation}\label{eq:<F0}
    \sum_{i\ge D}\sum_l l (\mathcal{O}(t)|\PP_{il} |\mathcal{O}(t))\le F_0b^{-D}\mathrm{e}^{\frac{t}{2}\left[c\kappa'+\kappa(b+\frac{1}{b})\right]},
\end{equation}
with constant $F_0, c$, for any $b>1$. Here we achieve $c=1.09$, but speculate that $c$ (along with the prefactor of the $\kappa$ term) may be further optimized by more sophisticated techniques within this quantum walk formalism. 

The proof starts from defining $F:=(\mathcal{O}| \mathcal{F} |\mathcal{O})$, where 
\begin{equation}
    \mathcal{F}:=\sum_{il} b^{i}l\PP_{il}.
\end{equation} 
Note that one can use $l^\nu$ in this definition instead of $l$, and the following still hold after generalization. However, we focus on $\nu=1$ because it turns out to give an optimal bound for butterfly velocity.
The goal of the proof is to bound $\frac{\mathrm{d}F}{\mathrm{d}t}$ by some coefficient times $F$ itself:
\begin{align}\label{eq:dFdtB}
    \frac{\mathrm{d}F}{\mathrm{d}t} &=(\mathcal{O}| [\mathcal{F},\LL] |\mathcal{O}) = \sum_i (\mathcal{O}| [b^{i}\sum_l l\PP_{il},\LL_{ii}] |\mathcal{O}) \nonumber\\ &+ \sum_{i} (\mathcal{O}| [b^{i}\sum_l l\PP_{il} + b^{j}\sum_l l\PP_{jl},\LL_{ij}] |\mathcal{O}) \nonumber\\
    &= \sum_i b^{i} G_{ii} + \sum_{i} (b^{i}G_{ij}+b^{j} G_{ji}),
\end{align}
where
\begin{align}
    G_{ii} &:=-2\sum_l l (\mathcal{O}|\LL_{ii}|\mathcal{O}_{il}) \nonumber\\ &=-2\sum_l l (\mathcal{O}_{il+1}+\mathcal{O}_{il-1}|\LL_{ii}|\mathcal{O}_{il}) \nonumber\\ &= 2\sum_{l>0}(\mathcal{O}_{il+1}|\LL_{ii}|\mathcal{O}_{il}),
\end{align}
where $|\mathcal{O}_{il}):=\PP_{il}|\mathcal{O})$. $G_{ij}$ is defined similarly: Let $\mathcal{O}_{il,jl'} = \PP_{il}\PP_{jl'}\mathcal{O}$. Since $(\mathcal{O}_{il}|\LL_{ij}|\mathcal{O}_{il})=0$, only the $\PP_{il\pm1}$ terms in e.g. $[Z_iZ_j,T_i^{lm}T_j^{l'm'}]=(Z_iT_i^{lm})m'T_j^{l'm'} + mT_i^{lm}(T_j^{l'm'}Z_j)$ contribute to the sum:
\begin{align}
    G_{ij} &:=-2\sum_l l (\mathcal{O}|\LL_{ij}|\mathcal{O}_{il})=2\sum_{l} (\mathcal{O}_{il+1}|\LL_{ij}|\mathcal{O}_{il}) \nonumber\\ &= 2\sum_{l}\sum_{l'>0} (\mathcal{O}_{il+1,jl'}|\LL_{ij}|\mathcal{O}_{il,jl'}).
\end{align}

We first bound $G_{ii}$ using eq.(\ref{eq:Kl<}):
\begin{align}
    G_{ii} &\le 2\sum_{l>0} \norm{\mathcal{O}_{il+1}}\norm{\PP_{il+1}\LL_{ii}|\mathcal{O}_{il})} \nonumber\\ &\le \frac{\kappa'}{2}\sum_{l>0}g(l)(l+1)\varphi_{il}\varphi_{il+1},\nonumber\\ &\le \frac{\kappa'}{4}\sum_{l>0}(l+1)^2\left[\frac{1}{l+1/2}\varphi_{il+1}^2+ \frac{1}{l+3/2}\varphi_{il}^2\right] \nonumber\\ &\le c\frac{\kappa'}{2}\sum_l l\varphi_{il}^2.
\end{align}
where $\varphi_{il}\equiv \norm{\mathcal{O}_{il}}$, and $g(l)\equiv \frac{(l+1)}{\sqrt{(l+1/2)(l+3/2)}}$ for short. 
Note that here the prefactor $c=\frac{55}{42}$, but it can be tightened to $1.09$ using a similar $\xi_l$ trick below, and keeping the first equation in eq.(\ref{eq:Kl<}), instead of the untight bound therein. 

To bound $G_{ij}$, we have
\begin{align}
    &\norm{\PP_{il+1}\LL_{ij}|\mathcal{O}_{il,jl'})}\nonumber\\ &\le\frac{1}{2}\frac{\kappa}{2S+1}\norm{\PP_{il+1} \sum_{mm'}\mathcal{O}_{mm'}\{Z_i, T^{lm}_i\} m'T^{l'm'}_j} \nonumber\\ &\le \frac{1}{2}\frac{\kappa}{2S+1}\varphi_{il,jl'} 2\sqrt{\frac{(l+1)^2-0^2}{(2l+1)(2l+3)}}(S+\frac{1}{2})l' \nonumber\\ &\le \frac{1}{4}\kappa g(l) l'\varphi_{il,jl'},
\end{align}
where we used expansion $\mathcal{O}_{il,jl'}=\sum_{mm'}\mathcal{O}_{mm'} T^{lm}_i T^{l'm'}_j$  where operator $\mathcal{O}_{mm'}$ is proportional to the identity on sites $i$ and $j$, and $a_{mm^\prime} := \sqrt{(\mathcal{O}_{mm'}|\mathcal{O}_{mm'})} $.  Since $\mathcal{O}_{mm^\prime}T^{lm}_i T^{l'm'}_j$ are orthogonal for different pairs of $(lm, l'm')$,  the norm is maximized simply by $a_{mm'}\propto\delta_{m0}\delta_{m'l'}$, as shown in the third line above. Then
\begin{align}\label{eq:Gijbound}
    G_{ij}&\le 2\sum_{l,l'} \norm{\mathcal{O}_{il+1,jl'}}\norm{\PP_{il+1}\LL_{ij}|\mathcal{O}_{il,jl'})} \nonumber\\ &\le \frac{\kappa}{2}\sum_{l,l'}l'g(l) \varphi_{il+1,jl'}\varphi_{il,jl'}\nonumber\\ &\le\frac{\kappa}{4} \sum_{l,l'}l' [(\xi_{l})^{-1}\varphi_{il+1,jl'}^2+ \xi_{l}g^2(l)\varphi_{il,jl'}^2]\nonumber\\ &= \frac{\kappa}{4} \sum_{l,l'} \varphi_{il,jl'}^2 l'[(\xi_{l-1})^{-1}+ \xi_{l}g^2(l)] \nonumber\\ &\le \frac{\kappa}{2} \sum_{l,l'} \varphi_{il,jl'}^2 l' = \frac{\kappa}{2} \sum_{l'} \varphi_{jl'}^2 l'.
\end{align}
In the third inequality we used $2g(l)\varphi_{il+1,jl'}\varphi_{il,jl'}\le (\xi_{l})^{-1}\varphi_{il+1,jl'}^2+ \xi_{l}g^2(l)\varphi_{il,jl'}^2$, and the last inequality holds as long as the positive sequence $\{\xi_l\}^\infty_{l=0}$ satisfies $(\xi_{l-1})^{-1}+ \xi_{l}g^2(l)\le 2$ with $(\xi_{-1})^{-1}=0$. This is true by setting $\xi_0=2/g^2(0)=3/2$ and iterating $\xi_l=[2-(\xi_{l-1})^{-1}]/g^2(l)$, which is solved by $\xi_l = \frac{2l+3}{2l+2}$. One can treat $G_{ji}$ similarly. Combining these into eq.(\ref{eq:dFdtB}),
\begin{align}
    \frac{\mathrm{d}F}{\mathrm{d}t} & \le c\frac{\kappa'}{2} \sum_{il} b^{i}l\varphi_{il}^2 + \frac{\kappa}{2}\sum_i b^i \left(\sum_{l'} \varphi_{jl'}^2 l'+b\sum_{l} \varphi_{il}^2 l\right)\nonumber\\ &= \frac{1}{2}\left[c\kappa'+\kappa(b+\frac{1}{b})\right]F,
\end{align}
Exponentiating this and using Markov's inequality finishes the proof of eq.(\ref{eq:<F0}).

\comment{
Defining the size vector $L=(L_1,\cdots,L_N)$ and $\varphi_L = \norm{\mathbb{P}_{1L_1}\cdots \mathbb{P}_{NL_N}|\mathcal{O}) }$, the last term can be written as
\begin{align}
    9\kappa \sum_L \varphi_L^2 \sum_i b^i(1+b)L_i L_{i+1}\le 9\kappa \sum_L \varphi_L^2 (\sqrt{b}+\frac{1}{\sqrt{b}})\sum_i b^i L_i^2= 9\kappa (\sqrt{b}+\frac{1}{\sqrt{b}}) F.
\end{align}
Thus, 
\begin{equation}\label{eq:dFnew}
    \frac{\mathrm{d}F}{\mathrm{d}t} \le \mu_b F + 9(\frac{17}{28}+c\kappa(b+\frac{1}{b}))F', \quad \mu_b\equiv 9+9\kappa(\sqrt{b}+\frac{1}{\sqrt{b}}),
\end{equation}
and
\begin{equation}
    F\lesssim F_0 \mathrm{e}^{\mu_b t}, \;\;\;\;\; \text{ where }F_0=\sum_ll^2 (\mathcal{O}|\mathbb{P}_{1l}|\mathcal{O})
    \end{equation}
if we assume $F'\ll F$ at large $t$. Alternatively, without this assumption, one can derive 
\begin{align}
    \frac{\mathrm{d}F'}{\mathrm{d}t} \le \frac{36}{\sqrt{15}}(\frac{1}{2}+\frac{\kappa}{2}(b+\frac{1}{b}))F'
\end{align}
using the same method and simply setting $\frac{l+1}{\sqrt{(l+1/2)(l+3/2)}}\le \frac{4}{\sqrt{15}}, (l\ge 1)$. Combining linearly with eq.(\ref{eq:dFnew}) gives the same result.
}

Since OTOC is upperbounded by the lhs of eq.(\ref{eq:<F0}) up to powers of $S$, we get a bound on the butterfly velocity:
\begin{equation}\label{eq:1dvB}
    v_{\mathrm{B}}\le \inf_{b>1} \frac{c\kappa'+\kappa(b+\frac{1}{b})}{2\ln b},
\end{equation}
which is independent on $S$.
\comment{The best Lieb-Robinson-type result we found, following \cite{PRXQuantum.1.010303}, is of the same form as (\ref{eq:1dvB}), but with $c=2$ (see Appendix \ref{app:LRB}). Thus the quantum walk approach not only improves $\lambda_{\mathrm{OTOC}}$ by a factor of $2$, but also gives a tighter constraint on the speed of information spreading in generic models.} 
Our proofs can easily be generalized to higher dimensional lattices (or even arbitrary graphs), as well as higher order interactions.

One might naively think that it is not possible to send signals or entanglement with a velocity too much faster than the butterfly velocity that characterizes chaos. However, for small system sizes, this is not the case. Following \cite{Tran:2020xpc, Kuwahara:2019rlw}, consider the two site Hamiltonian $H = S^{-1}(S+Z_1)Z_2$, and consider preparing an initial quantum state $|\psi(0)\rangle = \frac{1}{2}(|S\rangle + |-S\rangle)\otimes (|S\rangle + |-S\rangle)$.  Evolving for time $\tau = \pi/4S$, we find the entangled state $|\psi(\tau)\rangle = |S\rangle (|S\rangle+|-S\rangle)/2 - \mathrm{i} |-S\rangle (|S\rangle - |-S\rangle)/2 $.    Clearly, these two bits share one bit of entanglement.  For dynamics restricted to the $|\pm S\rangle$ states, this is the best possible transmission rate of entanglement \cite{karel}.  Hence, the generation of entanglement can proceed along much faster lines than the growth of operators, at least at early times.  This is another appearance of a ``hierarchy" of speed limits on different kinds of quantum information dynamics \cite{Tran:2020xpc}.

\section{Lieb-Robinson-type method}\label{app:LRB}
An alternative approach to bounding OTOCs in large-$S$ models was presented in \cite{PRXQuantum.1.010303}.  There, the authors proved bounds on $\lVert [S^\alpha_i(t),S^\beta_j]\rVert$, where $\lVert A \rVert$ denotes the maximal singular value of $A$.  To do this, and to get around the large operator norms of $\lVert S^\alpha \rVert = S$, the authors considered the large-$S$ Hilbert space to consist of the Dicke manifold of $2S$ interacting qubits with a permutation symmetric Hamiltonian, by writing  $2S^\alpha_i = \sum_{p=1}^{2S} \sigma^\alpha_{ip}$.  One can use a Lieb-Robinson bound for this enlarged system to bound the original problem, since we have $\norm{[S^\alpha_i(t), S^\beta_j]}\le \frac{1}{4}\sum_{pq} \norm{[\sigma^\alpha_{ip}(t), \sigma^\beta_{jq}]}$. 

\comment{
As noted in the main text, we consider the large-$S$ Hilbert space (on each site) to correspond to the Dicke manifold within an enlarged Hilbert space.  Namely, we write each large-$S$ spin as $2S$ spins-$1/2$: $2Z_i=\sum^{2S}_{p=1}\sigma^z_{ip}$, where $\sigma^z_{ip}$ is the Pauli $z$ matrix on the $p$-th spin-$1/2$ at site $i$, with $p=1,\ldots, 2S$.  We can use a Lieb-Robinson bound for this enlarged system to bound the original problem, since we have $\norm{[S^\alpha_i(t), S^\beta_j]}\le \frac{1}{4}\sum_{pq} \norm{[\sigma^\alpha_{ip}(t), \sigma^\beta_{jq}]}$, where $\norm{\cdot}$ is the conventional operator norm. Thus the Lieb-Robinson velocity of the enlarged system is also that of the original. This technique was developed in \cite{PRXQuantum.1.010303}. 
}

It is instructive to study the Lieb-Robinson bound for the spin chain; by simply turning off inter-site interactions, we also recover bounds for the kicked top arises as a simple case. For the spin-$1/2$ system, we use Corollary 7 of \cite{Chen:2019klo}:
\begin{equation}
    \norm{[\sigma^\alpha_{ip}(t), \sigma^\beta_{jq}]}\le 2\sqrt{3} \left[\exp(2|t|h)\right]_{ip,jq},
\end{equation}
where the $p$ indices run from $p=1,\ldots,2S$ and denote the auxiliary spin-$\frac{1}{2}$ degrees of freedom that we have added, and the matrix $h$ is given by
\begin{equation}
    h_{ip,iq} = \norm{\frac{\kappa'}{2S+1}\frac{2}{4}\sigma_{ip}\sigma_{iq}}(1-\delta_{pq}) \le \frac{\kappa'}{4S}(1-\delta_{pq}).
\end{equation}
Similarly
\begin{equation}
   h_{ip,(i+1)q}= \norm{\frac{\kappa}{2S+1}\frac{1}{4}\sigma_{ip}\sigma_{i+1,q}} \le\frac{\kappa}{8S}.
\end{equation}
 From the permutation symmetry of the $p$ index, we know that
\begin{align}
    (\mathrm{e}^{2th})_{ip, iq} &= u'_{ii}(t)\delta_{pq}+ u_{ii}(t)(1-\delta_{pq}), \nonumber\\ (\mathrm{e}^{2th})_{ip, j q} &= u_{ij}(t), \quad (j\neq i),
\end{align}
where
\begin{align}
    \dot{u'}_{ii} &= 2(2S-1)\frac{\kappa'}{4S}u_{ii} + 2(2S)\frac{\kappa}{8S}(u_{i,i-1}+u_{i,i+1}) \nonumber\\ &\le \kappa' u_{ii} + \frac{\kappa}{2}(u_{i,i-1}+u_{i,i+1}), \nonumber\\
    \dot{u}_{ii} & \le \frac{\kappa'}{2S}u'_{ii}+ \kappa' u_{ii} + \frac{\kappa}{2}(u_{i,i-1}+u_{i,i+1}), \nonumber\\
    \dot{u}_{ij} &\le \kappa' u_{ij} + \frac{\kappa}{2}(u_{i-1,j}+u_{i+1,j}).
\end{align}
At zeroth order in $1/S$, we can set $u'_{ii}\equiv 0$, so that matrix $u=\mathrm{e}^{2th}$ where $2h_{ij}=\kappa'\delta_{ij}+\frac{\kappa}{2}(\delta_{i,j+1}+\delta_{i,j-1})$. When there is only one site $i$, we get $\lambda_{\mathrm{OTOC}}\le 4h_{ii}= 2\kappa'$ for the kicked top. Assuming periodic boundary condition of the chain and $L\rightarrow \infty$,
\begin{equation}
    u_{i,i+x}(t)\propto \mathrm{e}^{\kappa't} J_x(-\mathrm{i}\kappa t) = \mathrm{e}^{\kappa'x/v} J_x(-\mathrm{i}\kappa x/v),
\end{equation}
where we parametrize $t=x/v$. Uniform asymptotic expansion of the Bessel function at large $x$ shows that the critical $v=v_{\mathrm{LR}}$ is given by
\begin{equation}
    \kappa'/\kappa=v'\ln(v'+\sqrt{1+v'^2}) - \sqrt{1+v'^2},
\end{equation}
where $v'=v_{\mathrm{LR}}/\kappa$. This is equivalent to (\ref{eq:1dvB}) with $c=2$.

When studying either the kicked top or the spin chain, the additional factor of 2 noted above arises because Lieb-Robinson bounds are for commutators: $\lVert [S^\alpha(t),S^\beta]\rVert \lesssim S\mathrm{e}^{\kappa t}$, while the OTOC involves a \emph{squared commutator}, which grows twice as fast. The quantum walk methods are more effective at accounting for the destructive interference between growing operators, and thus improve upon the best known Lieb-Robinson-type bounds by a factor of 2.

\section{Towards Holographic Models}
One of the motivations for this work was also to understand the similarities and differences between operator growth in holographic models like the Sachdev-Ye-Kitaev (SYK) model, and large-$S$ coupled spin dynamics.  A ``hybrid" model is 
\begin{equation}
    H = \frac{\kappa}{S} Z_i Z_i + \frac{1}{S\sqrt{N}} J_{ij}  S^\alpha_i S^\alpha_j,
\end{equation} 
with $J_{ij}$ standard Gaussian random variables.  See \cite{Scaffidi:2017ghs,Pappalardi:2019kkh} for qualitatively similar models.  We have used Einstein summation convention on indices.  A rigorous OTOC bound for such a model would be quite involved \cite{lucas2020non} as the disorder average is highly non-trivial. We postulate that a rigorous bound on $\lambda_{\mathrm{OTOC}}$ will reveal two contributions to the LE: one from on-site growth ($Z_i \rightarrow X_i^2$) and one from inter-site growth ($Z_i \rightarrow X_i X_j$).  More practically, we expect that the inter-site  dynamics is rather similar to operator growth in the SYK model \cite{Roberts:2018mnp,lucas2020non}, while the on-site dynamics could disrupt the constructive interference patterns which lead to exponential operator growth in SYK.  Unlike in the large-$S$ Hilbert space, typical operators of size $s$ in SYK grow at a rate $\sqrt{s}$ rather than $s$.  We are not sure whether or not this qualitative difference in operator growth is of much practical consequence, e.g. for experimentalists aiming to study quantum gravity via many-body chaos in the lab.

\section{Outlook}
We have proven reasonably sharp bounds on quantum OTOCs and operator growth in semiclassical spin chains and kicked top models.  Our results  improve upon Lieb-Robinson bounds and provide a useful new mathematical framework for the study of quantum information dynamics and quantum chaos in a semiclassical limit.  We hope that similar methods will also be useful in constraining quantum dynamics with bosonic degrees of freedom, another hard problem with large (infinite) dimensional Hilbert spaces where very few results are known \cite{Nachtergaele_2008,LRion}. Our methods may also lead to sharp answers to the question of which experimentally-realizable microscopic models can faithfully mimic the dynamics of holographic quantum gravity.

The semiclassical spin models we have studied here can also be interpreted in terms of many-body quantum mechanical models with $S$ spin-$\frac{1}{2}$ degrees of freedom, with SU(2)-symmmetric Hamiltonian.  When restricted to the Dicke manifold (total angular momentum $S(S+1)$), our bounds describe the resulting dynamics.  Remarkably, OTOC growth \emph{speeds up} quite subtantially in the Dicke manifold, relative to in typical states in Hilbert space, where with the normalization (\ref{eq:generalH}), $\lambda_{\mathrm{OTOC}}=0$ can be proved \cite{Yin:2020pjd}.  This is a notable counterexample to the generic expectation that quantum dynamics should slow down in constrained subspaces \cite{Han:2018bqy,Chen:2020bmq}.  We hope that our methods can be used to help understand the robustness of certain spin squeezed states to generic perturbations \cite{perlin2020spin}, which has recently been observed numerically and has broad applications to quantum metrology.

\section*{Acknowledgements}
We thank Brian Swingle and Zhiyuan Wang for useful discussions.  AL was supported by a Research Fellowship from the Alfred P. Sloan Foundation.

\onecolumngrid

\begin{appendix}
\section{Operator growth bounds beyond the kicked top}\label{app:proofs}
Consider the Hamiltonian (\ref{eq:generalH}).  In this appendix, we focus on models where if $h(n_x,n_y,n_z)\ne 0$, then $n_x+n_y+n_z \le n_0$.  Then, if $|h(\bm{n},t)|\le h(\bm{n})$, we will prove below that
\begin{equation}
    \lambda_{\mathrm{OTOC}}\le  2(n_0-1)\sum_{\bm{n}}h(\bm{n})\sqrt{\sum_\alpha n_\alpha -1}\sum_\alpha \frac{\sqrt{n_\alpha} (n_\alpha+1)!}{(\lfloor n_\alpha/3\rfloor!)^3}.  \label{eq:appthm1}
\end{equation}

Since $[H, T^{lm}]$ contains components of size $l'=l-n_0+1,\cdots,l+n_0-1$, the corresponding quantum walk equation is 
\begin{equation}
    \frac{\mathrm{d}\phi_l}{\mathrm{d}t} = \sum_{l'<l}K_{ll'}(t)\phi_{l'}(t) - \sum_{l'<l} K_{l'l}(t)\phi_{l'}(t),
\end{equation}
where $|K_{l'l}|\le \mathcal{K}_{l'l}=\norm{\mathbb{Q}_{l'}\LL\mathbb{Q}_l}$. We group neighboring $n_0-1$ sizes as a block labeled by $L$. Namely, block $L$ corresponds to a set $R_L$ of size, where
\begin{equation}
    R_L := \left\{\begin{array}{ll}
    \{1\} & L=1 \\
    \{l\in \mathbb{Z}:2+(n_0-1)(L-2)\le l\le \min\{ 1+(n_0-1)(L-1), 2S\}\}     & 2\le L\le L_m:=\left\lceil\frac{2S-1}{n_0-1}\right\rceil+1
    \end{array}\right.
\end{equation}
Define \begin{equation}
    \mathbb{Q}_L := \sum_{l\in R_L} \mathbb{Q}_l
\end{equation}and 
\begin{equation}
    \phi_L(t) := \norm{\mathbb{Q}_L|\mathcal{O}(t))}.
\end{equation} Similar to (\ref{eq:qwalk}), 
\begin{equation}
    \frac{\mathrm{d}\phi_L}{\mathrm{d}t} = K_{L-1}(t)\phi_{L-1}(t) - K_{L}(t)\phi_{L+1}(t),
\end{equation}
where
\begin{equation}
    |K_L|\le \mathcal{K}_L := \norm{\mathbb{Q}_{L+1}\LL\mathbb{Q}_L} = \max\slr{\max_{l\in R_L}\sum_{l'\in R_{L+1}} \mathcal{K}_{l'l}, \max_{l'\in R_{L+1}}\sum_{l\in R_L} \mathcal{K}_{l'l}}.
\end{equation}
Define $\tilde{F}:= \sum_L L(L+1)\phi_L^2$ so that $F\le (n_0-1)^2\tilde{F}$. If we can show that $\mathcal{K}_L\le \eta(L+1)$ with $\eta$ independent of $L,S$, then \begin{align}
    \frac{\mathrm{d}\tilde{F}}{\mathrm{d}t} &= 2\sum_L L(L+1)\phi_L \frac{\mathrm{d}\phi_L}{\mathrm{d}t} 
    \le \sum_L 2(L +1)\mathcal{K}_L \left(\phi_L^2 + \phi_{L+1}^2\right)\le 4\eta \sum_L [L(L+1)+\frac{1}{2}]\phi_L^2 = 4\eta\tilde{F}+ 2\eta,
\end{align}
which leads to $\lambda_{\mathrm{OTOC}}\le 4\eta$. 

Now, we find a bound for $\eta$. We  write the Hamiltonian as $H=\sum_{\bm{n}} H_{\bm{n}}$. Correspondingly, $\LL=\sum_{\bm{n}} \LL_{\bm{n}}$ and $\mathcal{K}_L=\sum_{\bm{n}} \mathcal{K}_{L,\bm{n}}$. For a given $\bm{n}$, $[X^{n_x}Y^{n_y}Z^{n_z},\mathcal{O}] = X^{n_x}Y^{n_y}[Z^{n_z},\mathcal{O}] + X^{n_x}[Y^{n_y},\mathcal{O}]Z^{n_z}+ [X^{n_x},\mathcal{O}] Y^{n_y}Z^{n_z}$. For the first term, because the maximal absolute eigenvalue of $X$ and $Y$ is $S$, we have \begin{align}
    \norm{ X^{n_x}Y^{n_y}[Z^{n_z},\mathcal{O}^l]} \le S^{n_x+n_y} \norm{[Z^{n_z},\mathcal{O}^l]}\le S^{n_x+n_y} \max_m \norm{[Z^{n_z},T^{lm}]},
\end{align}
where $\mathcal{O}^l$ is an arbitrary operator of size $l$, and we used the fact that $[Z^{n_z}, T^{lm}]$ are orthogonal for different $m$. Assume $m>0$ and let $n=n_z$ for a moment  for simplicity. Let 
\begin{equation}
    C_m(l,l')=2\sqrt{\pi/3} c(l,l')\langle 10lm|l'm\rangle = \left\{\begin{array}{ll}
    \sqrt{\frac{(l+1)^2-m^2}{(2l+1)(2l+3)}}\sqrt{(S+\frac{1}{2})^2-\frac{1}{4}(l+1)^2}     & l'=l+1 \\
    m/2     & l'=l \\
    \sqrt{\frac{l^2-m^2}{(2l-1)(2l+1)}}\sqrt{(S+\frac{1}{2})^2-\frac{1}{4}l^2}    & l'=l-1
    \end{array}\right.
\end{equation}
which makes $Z T^{lm}=\sum_{l'}C_m(l,l')T^{l'm}$ and $T^{lm}Z=\sum_{l'}(-)^{l'-l+1}C_m(l,l')T^{l'm}$. Furthermore, $C_m(l,l\pm1)\le (2S+1)/\sqrt{15}$ except $C_m(1,0)$, which does not enter in the following results. Then
\begin{align}\label{eq:sum_p}
    [Z^n,T^{lm}] = \mlr{Z^{n-1}\sum^{l+1}_{l''=l-1}C_m(l,l'')T^{l''m}-\sum^{l+1}_{l''=l-1}(-1)^{l''-l+1}C_m(l,l'')T^{l''m}Z^{n-1}}=\cdots= 2\sum'_p \mlr{\prod^n_{j=1}C_m(p_j)} T^{l'm} ,
\end{align}
Here array $p$ denotes a path in size space from $l$ to $l'$. For example, $p=(l+1,l+1,l,l+1)$ is a path $l\rightarrow l+1\rightarrow l+1\rightarrow l\rightarrow l+1$ with $l'=l+1$ and $n=4$.   Let $n_\pm$ be the number of terms in which $l \rightarrow l\pm 1$, and $n^\prime$ the number of steps in which $l\rightarrow l$. In (\ref{eq:sum_p}) we also denote $C_m(p_j)=C_m(p_{j-1},p_{j})$ where $p_0\equiv l$, and $\sum'_p$ only contains path $p$ with an odd $n'$. Suppose a path has $n_+$ steps of forward hopping (i.e. increasing $l$ by 1), and $n_-$ steps of backward hopping. They satisfy $n=n_++n_-+n', l'-l=n_+-n_-$. The number of paths with a given $n'$ will be $\le\frac{n!}{n_+!n_-!n'!}$, where the inequality is due to presence of boundaries $1\le l\le 2S$. For one such path, \begin{equation}
    \prod^n_{j=1}C_m(p_j)\le \left(\frac{m}{2}\right)^{n'}\left(\frac{2S+1}{\sqrt{15}}\right)^{n-n'},
\end{equation}
and therefore
\begin{align}\label{eq:QZT}
    \norm{\mathbb{Q}_{l'}[Z^n,T^{lm}]}&\le 2 \sum^{n}_{n'=1, 3,5,\ldots} \frac{n!}{n_+!n_-!n'!} \slr{\frac{\sqrt{15}m}{2(2S+1)}}^{n'}\slr{\frac{2S+1}{\sqrt{15}}}^{n} \nonumber\\ &\le 2 \frac{n+1}{2} \frac{n!}{(\lfloor n/3 \rfloor!)^3} \slr{\frac{\sqrt{15}S}{2S+1}}^{n-1}\frac{\sqrt{15}m}{2(2S+1)}\slr{\frac{2S+1}{\sqrt{15}}}^{n} \le \frac{l}{2} S^{n-1}  \frac{(n+1)!}{(\lfloor n/3\rfloor!)^3}.
\end{align}
Putting back $n_z=n$, $\norm{\mathbb{Q}_L |[Z^{n_z},T^{lm}])}$ is then bounded by adding a $\sqrt{n_z}$ factor, since there are $n_z$ choices of $l'$.  Note that the norm here represents the Frobenius norm, which is why we only require a square root here:  the operators with different $l^\prime$ are necessarily orthogonal.  Gathering all the prefactors,  \begin{equation}
    \norm{\LL_{\bm{n}}\mathbb{Q}_l}\le h(\bm{n}) \frac{l}{2}\sum_\alpha \frac{\sqrt{n_\alpha} (n_\alpha+1)!}{([n_\alpha/3]!)^3}.
\end{equation}
Finally, using that each $\mathbb{Q}_l$ projects onto a disjoint subspace, together with the Frobenius norm,
\begin{align}
    \mathcal{K}_{L,\bm{n}}&\le \sqrt{\sum_\alpha n_\alpha -1}\max\slr{ \max_{l\in R_L}\norm{\mathbb{Q}_{L+1}\LL_{\bm{n}}\mathbb{Q}_l},\max_{l\in R_{L+1}}\norm{\mathbb{Q}_{L}\LL_{\bm{n}}\mathbb{Q}_l} }\notag \\
    &\le h(\bm{n})\sqrt{\sum_\alpha n_\alpha -1} \frac{(n_0-1)(L+1)}{2}\sum_\alpha \frac{\sqrt{n_\alpha} (n_\alpha+1)!}{([n_\alpha/3]!)^3}.
\end{align}
which leads to (\ref{eq:appthm1}).

\end{appendix}

\bibliography{large_spin}

\end{document}